\documentclass[prl,twocolumn,amsmath,amssymb,floatfix]{revtex4}
\usepackage[pdftex]{graphicx}
\usepackage{subfigure}
\usepackage[pdftex,debug=true,breaklinks=true,colorlinks=true]{hyperref}
\usepackage[ansinew]{inputenc}
\usepackage{color}
\usepackage{bm} % boldmath
\newcommand{\bra}[1]{\langle #1|}
\newcommand{\ket}[1]{|#1\rangle}
\newcommand\sx{\raisebox{-1.75pt}{\resizebox{.015\textwidth}{!}{\includegraphics{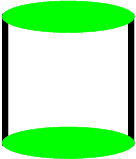}}}}
\newcommand\sy{\raisebox{-1pt}{\resizebox{.017\textwidth}{!}{\includegraphics{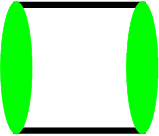}}}}
\newcommand\sxy{\raisebox{-1pt}{\resizebox{.015\textwidth}{!}{\includegraphics{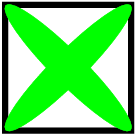}}}}
\newcommand\txy{\raisebox{-1pt}{\resizebox{.015\textwidth}{!}{\includegraphics{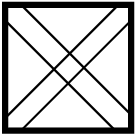}}}}
\newcommand\neel{\mbox{N\'eel\ }}
\newcommand\ie{\mbox{{\it i.e.}\ }}
\newcommand\SUdeux{\mbox{$SU(2)$\ }}
\newcommand\jj{\mbox{$J_1-J_2$\ }}
\newcommand\Himp{\mbox{$\mathcal{H}|_\square\ $}}
\newcommand\up{\uparrow}
\newcommand\down{\downarrow}
\newcommand\frust{\mbox{$t_2/t_1$\ }}
\newcommand\pp{\mbox{$(\pi,\pi)$\ }}
\newcommand\zp{\mbox{$(0,\pi)$\ }}
\newcommand\pz{\mbox{$(\pi,0)$\ }}
\def\lsim{\mathrel{\rlap{\lower3.5pt\hbox{\hskip.5pt$\sim$}}
\raise1pt\hbox{$<$}}}                % less than or approx. symbol
\def\gsim{\mathrel{\rlap{\lower3.5pt\hbox{\hskip.5pt$\sim$}}
\raise1pt\hbox{$>$}}}                % greater than or approx.symbol
\begin{document}
\title{Parent phases of doped Mott insulators: a Cluster DMFT study}
\author{J.-C.~Domenge}
\email{domenge@physics.rutgers.edu}
\affiliation{Department of Physics and Astronomy and \mbox{Center for
Condensed Matter Theory},\ Rutgers University,\ Piscataway,\ NJ
08854-8019}
\author{G.~Kotliar}
\affiliation{Department of Physics and Astronomy and \mbox{Center for
Condensed Matter Theory},\ Rutgers University,\ Piscataway,\ NJ
08854-8019}
\date{\today} % It is always \today, today, but any date may be
%explicitly specified
%
\begin{abstract}
We investigate the insulating phases of a frustrated Hubbard model in
its strong coupling regime at half-filling. We pay special attention to
all the symmetry breaking instabilities that can be described by
Dynamical Mean Field Theory (DMFT)  on a square plaquette. We identify
several mean-field solutions, two \neel states breaking the \SUdeux
symmetry,  a dimer phase and  a Mott insulator which doesn't break any
symmetry. The singlet to singlet fluctuations soften dramatically in
the latter phase, giving rise to dimerization fluctuations as well as chiral
fluctuations that are both low-lying. We present a simple picture of the
different DMFT states and their evolution with frustration.
\end{abstract}
\maketitle
Frustrated two dimensional Mott insulators are interesting in their own
right, and are realized in Helium films as well as in transition metal
and  organic compounds. They received enormous attention, following the
discovery of  superconductivity in copper oxide based materials  and the
suggestion~\cite{a87} that the phenomenon of high temperature
superconductivity is connected to the doping of a "spin liquid", namely
a Mott insulator without magnetic long range order at zero temperature.
% there has been intensive efforts to understand how
%high temperature superconductivity can emerge from the doping of a Mott
%insulator. In an ideal treatment of high-temperature
%^superconductivity, In this line of thought,

In this line of thought one would  like to understand spin liquids,
metals and their relation to superconducting states, within one same
framework. Many  insights into this problem were obtained using mean
field methods such as slave boson mean field theory and  the
\mbox{large-N} expansion.~\cite{sr91}
% underlying normal state with no broken symmetry and
%monitor the emergence upon doping of the gauge symmetry breaking in the
%electronic structure. However, tracing such a path between the
%superconductor and its normal parent is difficult both experimentally
%and theoretically, since some form of magnetic order usually intervenes
%at small doping.  In mean-field theories the symmetry of the solution
%may be fixed a priori as part of the mean-field ansatz. Hence it is
%always possible to enforce a totally symmetric normal state and monitor
%its evolution into a superconductor upon doping.  Implementations of
%this idea, using slave bosons or Guztwiller mean field theory, were
%very successful, for instance in predicting  the symmetry of the
%pairing order parameter in cuprate superconductors (for a review see
%for instance Ref.~\onlinecite{lnw06}).
More recently the development of Dynamical Mean Field Theory (DMFT) and
its cluster extensions~\cite{gkk96,ksh06,mjp05,tks06} have  improved the
description of the unusual electronic structure of strongly correlated
materials. Indeed this approach applied to simple models accounts for
many properties of both the \mbox{$d$-wave} superconducting state and
the normal state of copper oxides~\cite{cck05,sk06} and
\mbox{$\kappa$-organics}.~\cite{kt06}
% in all different
%implementations of cluster DMFT on a $2\times2$ plaquette. The latter
%also accounts for the formation of the temperature-dependant Fermi arcs
%observed in photoemission experiments on cuprates,~ \cite{cck05,sk06}
%as well as the correlation strength and the frustration dependance of
%superconductivity in $\kappa$-organics.~ \cite{kt06}.

%THIS IS NOT CORRECT. TREMBLAY DID THE ORGANIC. FOR AN EARLY STUDY OF
%SPIN LIQUIDS IN CLUSTER DMFT SEE.  KAGOME LATTICE.  DMFT can also be
%used to explore broken symmetry states. In Ref.~\onlinecite{kt06} the
%usual hopping between nearest neighbors (NN) was supplemented with a
%frustrating hopping between next-nearest neighbors (NNN) and the
%frustration at which the \neel long-range order ceases to exist was
%identified.  However, other competing orders, such as dimerized phases,
%were not investigated although they are very likely to develop in such
%(effective) frustrated Heisenberg systems. STRESS WE DO t-t' MODEL ?
In this letter, we make a general mean-field ansatz to study
symmetry-breaking solutions describing the insulating phases of the
half-filled \mbox{$t_1-t_2$} Hubbard model at large $U$, using cluster
DMFT on a $2\times2$ plaquette. We identify several Mott insulating
phases and we take advantage of the substantial simplifications provided
by the CDMFT method to obtain a simple understanding of their local
properties. Such a program was carried out for the anisotropic triangular
lattice in Ref.~\onlinecite{kt06}, while in
Refs.~\onlinecite{gsf01,okt06} DMFT methods were applied to the Mott
insulating regime on the kagom\'e lattice. Unlike the above studies,
however, our DMFT ansatz is compatible with dimerization, a common
instability in frustrated spin systems.
%Our aim is to locate a region where the mean-field insulator is indeed
%a fully symmetric paramagnet and investigate its properties.

We start with a frustrated Hubbard model on the square lattice
\begin{equation}
\mathcal{H}=-\displaystyle{\sum_{(i,j),\sigma}}(t_{ij}c_{i\sigma}^{\dagger}c_{j\sigma}+h.c.)+U\displaystyle{\sum_{i}}n_{i\up}n_{i\down}-\mu\displaystyle{\sum_{i,\sigma}}n_{i\sigma}
\label{eqn:H}
\end{equation}
where $c_{i\sigma}$ (resp. $c_{i\sigma}^{\dagger}$) destroys (resp.
creates) an electron at site $i$ with spin projection
\mbox{$\sigma$} on the $z$-axis,
\mbox{$t_{ij}=t_1$} \mbox{(resp. $t_2$)} for pairs $(i,j)$ of NN
\mbox{(resp. NNN)}, and \mbox{$t_{ij}=0$} otherwise.
$U$ is the onsite Coulomb repulsion and $\mu$ is the chemical potential.

We investigate the insulating phases of~(\ref{eqn:H}) at half-filling
using Cellular DMFT on a $2\times2$ plaquette. To be more specific, we
consider the following Anderson Impurity Model
\begin{eqnarray}
\mathcal{H}_{\text{AIM}}=\mathcal{H}|_\square\nonumber&+&\displaystyle{\sum_{(\alpha,\beta),\sigma}}(\epsilon_{\alpha\beta,\sigma}a_{\alpha\sigma}^{\dagger}a_{\beta\sigma}+h.c.)\nonumber\\
&+&\displaystyle{\sum_{(i,\alpha),\sigma}}(V_{i\alpha,\sigma}c_{i\sigma}^{\dagger}a_{\alpha\sigma}+h.c.)
\label{eqn:HAIM}
\end{eqnarray}
where \Himp is the restriction of the original Hamiltonian~(\ref{eqn:H})
to one plaquette $\square$ (the impurity), and the last two terms read
the kinetic energy of the conduction bath and the hybridization between
the bath and the plaquette, respectively. In substance, solving the
self-consistent Cellular DMFT equations amounts to finding a set of bath
parameters
$\{\epsilon_{\alpha\beta,\sigma},V_{i\alpha,\sigma}\}_{\alpha,\beta,\sigma}$
that mimic the influence of the infinite square lattice surrounding
plaquette $\square$ in the original model~(\ref{eqn:H}).~\cite{gkk96}

On the technical side, the use of Exact Diagonalization to obtain the
ground state of~(\ref{eqn:HAIM}) severely constrains the size of the
bath (8 sites in the present study).  Further, numerical tractability
requires that the total number of bath parameters be even more limited,
usually by imposing additional symmetry relations among them. This step
requires extra care, especially when investigating possible
symmetry-breaking instabilities. Indeed, there is no {\it spontaneous}
symmetry breaking in a finite system, hence the symmetry of the ground
state of~(\ref{eqn:HAIM}) is merely that of the Hamiltonian itself.
Finally, attention must also be paid to the CDMFT self-consistency
condition. The latter describes the embedding of the impurity plaquette
$\square$ in the infinite lattice, which must be compatible with the
symmetry of the thermodynamic ground state. These points will be
discussed at greater length elsewhere.~\cite{dk08b}

We first present the mean-field ground states of~(\ref{eqn:H}) at
half-filling and for large $U$: in the following \mbox{$t_1=1$} and
\mbox{$U=16$}. We have checked that this $U$ puts the system well within
the Mott insulating phase. In this regime the cost of double occupancy
is avoided through the coherent hopping of two or more electrons, and to
lowest order in $t_{1,2}/U$,~(\ref{eqn:H}) reduces to the \jj Heisenberg
model, with \mbox{$J_i=4t_i^2/U(1+\mathcal{O}((t/U)^2))$}.~\cite{dk08b}
Hence the meanigful frustration parameter is $(t_2/t_1)^2$. For finite
$U$, higher order processes also contribute and generate longer exchange
loops, starting with four-spins exchange at order $t_i^4/U^3$.

\begin{figure}[h]
\resizebox{.5\textwidth}{!}{\includegraphics[viewport=45 35 780
525,clip]{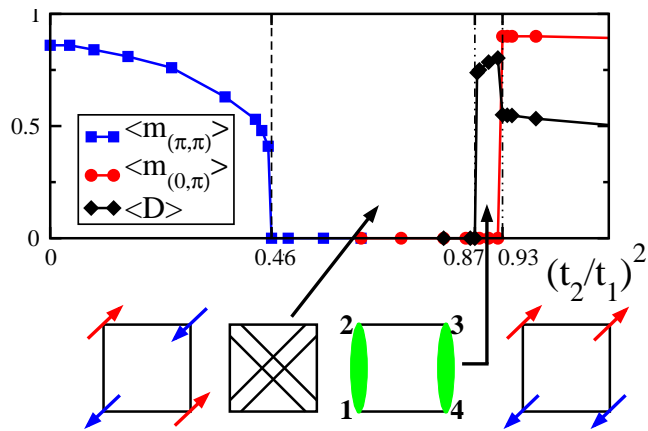}}
\caption{(Color online) Evolution of the \neel and dimer order
parameters computed on the impurity plaquette for increasing frustration
\frust (see text). Cartoon pictures of the various phases on the
impurity plaquette are shown: in both \neel phases the relative
orientation of the spins is indicated, while in the paramagnetic phases,
triplets and singlets are pictured with parallel lines and green
ellipses, respectively.}
\label{fig:OP}
\end{figure}

Figure~\ref{fig:OP} shows the evolution of three order parameters on the
impurity upon increasing the frustration ratio \frust. First, the usual
\neel order is detected by computing the normalized staggered
magnetization \mbox{$\langle
m_{\bm{k}_\text{I}}\rangle=\frac{2}{N_c}\sum_{i\in\square}e^{j\bm{k}_\text{I}\cdot\bm{r}_i}\langle
S_i^z\rangle$}, where \mbox{$\bm{k}_\text{I}=(\pi,\pi)$} and
\mbox{$N_c=4$}.  Another \neel order is anticipated at large frustration
from previous studies of the closely-related \jj model (see
Ref.~\onlinecite{ml05} and references therein). It corresponds to the
entropic selection of one of the two degenerate impurity momenta
\mbox{$\bm{k}_\text{I}=(0,\pi)$} and \pz.  Finally we compute
\mbox{$\langle
D\rangle=\langle\bm{S}_2\cdot\bm{S}_3-\bm{S}_1\cdot\bm{S}_2\rangle$}
which evidences the $x-y$ symmetry breaking of the spin correlations.
Note that $\langle D\rangle$ is non-zero if the impurity plaquette
carries singlets on parallel NN bonds, but also for the \zp and \pz
\neel orders. Hence the dimerization itself is detected through both i)
\mbox{$\langle D\rangle\neq0$} and ii) \mbox{$\langle
m_{(0,\pi)}\rangle=\langle m_{(\pi,0)}\rangle=0$}.

At \mbox{$t_2=0$} we recover the \pp \neel order with a staggered
magnetization renormalized to about 86\% of its classical value.  Upon
increasing the frustration, $\langle m_{(\pi,\pi)}\rangle$ is increasingly
renormalized until quantum fluctuations completely wipe out the \neel
order at \mbox{$(t_2/t_1)^2\simeq0.46$}.

For \mbox{$0.46\lsim(t_2/t_1)^2\lsim0.87$} the ground state is a singlet
formed by the four-spins of the impurity. Since it doesn't break the
point symmetry of a square, we will refer to it as a fully symmetric
paramagnet in the following.  For \mbox{$0.87\lsim(t_2/t_1)^2\lsim0.93$}
the ground state is still paramagnetic (all \neel order parameters are
zero) but it spontaneously breaks the rotational symmetry of the lattice
and dimerizes in an almost pure product of singlets on the plaquette, as
evidenced by $D\simeq3/4$.

Finally \SUdeux spontaneously breaks again for
\mbox{$(t_2/t_1)^2\gsim0.93$} where the \zp \neel phase kicks in through
a first-order transition, as suggested by the abrupt increase of
$\langle m_{(0,\pi)}\rangle$ jumping from 0 to about 90\% of its
classical value. This is further supported by the simultaneous existence
of two solutions to the CDMFT equations in this region, namely the fully
symmetric paramagnet and the \zp \neel order.~\cite{dk08b}
%When decreasing the frustration \frust from the \zp \neel phase, the
%\neel solution can be continued down to \mbox{$(t_2/t_1)^2<0.93$},
%evidencing the existence of two solutions of the DMFT equations. This
%further supports the scenario of a first-order transition and will be
%discussed in greater detail elsewhere.~\cite{dk08b}}

{\it Interpretation} Cellular DMFT provides a local picture of each of
the phases of model~(\ref{eqn:H}), in terms of the states of the
self-consisent impurity model~(\ref{eqn:HAIM}).

In semi-classical phases, such as the \pp \neel phase at small $t_2$,
tracing out the bath degrees of freedom results in a reduced density
matrix for the impurity plaquette that has most of its weight in one
state of the plaquette. This state resembles a classical,
antiferromagnetic, spin configuration dressed by quantum fluctuations.
The orientation of the spin is picked up by the self-consistent Weiss
field.

To gain more insight into the fully symmetric phase that takes over at
intermediate frustration \mbox{$0.46\lsim(t_2/t_1)^2\lsim0.87$}, we
investigate the nature of the {\it two} lowest eigenstates of the
impurity model, noted $\ket{0}$ and $\ket{1}$.  From each one of these
states we construct the reduced ''density matrix'' of the impurity,
defined by
\mbox{$\rho_\text{I}^\alpha=\text{Tr}_\text{B}\ket{\alpha}\bra{\alpha}$}
where \mbox{$\alpha=0,1$} and $\text{Tr}_\text{B}$ denotes the partial
trace over the bath degrees of freedom. The spectral decomposition of
the operators $\rho_\text{I}^\alpha$ reveals that most of the weight
($\sim95\%$) is carried by a single state of the plaquette noted
$\ket{\alpha}_\text{I}$, {\it i.e.}
\mbox{$\rho_\text{I}^\alpha\simeq\ket{\alpha}_\text{I}\cdot{}_\text{I}\bra{\alpha}$}.
This means that
\mbox{$\ket{\alpha}\simeq\ket{\alpha}_\text{I}\ket{\alpha}_\text{B}$},
\ie the impurity and the bath are only weakly entangled in the subspace
spanned by $\ket{0}$ and $\ket{1}$, at stark contrast with the situation
in single-site DMFT.

Further, the nature of the impurity states $\ket{\alpha}_\text{I}$
provide a simple mean-field picture of the paramagnetic phases and 
their evolution with $t_2/t_1$. Namely, in the fully symmetric phase,
$\ket{0}_\text{I}$ is  the singlet obtained by adding two triplets along
the diagonals $(1,3)$ and $(2,4)$ of the plaquette, noted $\ket{\txy}$,
only lightly dressed by quantum fluctuations. Similarly,
$\ket{1}_\text{I}$ is the product of two diagonal singlets, noted
$\ket{\sxy}$.

We will show that the disappearance of the fully symmetric paramagnet at
\mbox{$(t_2/t_1)^2\simeq0.87$} is driven by the closing of the gap
between these two singlets. We believe that this is the {\it local}
mean-field description of the abundance of low-lying singlets commonly
observed at Quantum Phases Transition points in frustrated quantum
magnets.~\cite{ml05} This is supported by the observation that the two
impurity singlets are orthogonal $\bra{\sxy}\txy\rangle=0$, and span
exactly the two-dimensional subspace of the half-filled impurity with
total spin $S_\text{I}=0$. Hence, within plaquette DMFT there is simply
no additional, linearly independant, half-filled impurity singlet
susceptible to condense.

When the singlet gap closes, the zero-energy subspace is enlarged to
$\{\ket{\sxy},\ket{\txy}\}$. Increasing the frustration in
\mbox{$0.87\lsim(t_2/t_1)^2\lsim0.93$} lifts this degeneracy and selects
a particular direction within this subspace, either
\mbox{$\ket{\sy}=\sqrt{3}/2\ket{\txy}+1/2\ket{\sxy}$}, or
\mbox{$\ket{\sx}=\sqrt{3}/2\ket{\txy}-1/2\ket{\sxy}$}.

{\it Dynamical singlet correlations} More insight into the paramagnetic
phases can be gained by considering higher-order spin correlations.
Indeed, in a seminal paper~\cite{wwz89} Wen discussed a particular class
of spin liquids, breaking both Parity (P) and Time reversal (T), whose
order parameter is the scalar chirality \mbox{$\langle
E_{ijk}\rangle=\langle\bm{S}_i\cdot\bm{S}_j\times\bm{S}_k\rangle$}.  The
relevance of these chiral spin liquids for the \jj model was claimed by
the author, on the premise that the coherence in the collective dynamics
of three spins on a plaquette is generally enhanced by a non-zero $J_2$,
possibly to the point where $\langle E_{ijk}\rangle\neq0$.  Beyond the
original mean-field study by Wen, an early numerical work of the \jj
model evidenced the softening of chiral fluctuations, although no
long-range chiral order was found.~\cite{pgb91} Within the CDMFT
treatment of the related Hubbard model~(\ref{eqn:H}), the weak
entanglement between the bath and the impurity  yields
\mbox{$\bra{0}E_{ijk}\ket{0}\simeq\bra{\txy}E_{ijk}\ket{\txy}$} in the
fully symmetric phase.  Consider for instance the triplet $(1,2,3)$ of
impurity sites. Direct computation shows that
\mbox{$E_{123}\ket{\txy}=i\sqrt{3}/4\ket{\sxy}$} and
\mbox{$E_{123}\ket{\sxy}=-i\sqrt{3}/4\ket{\txy}$}. Hence
\mbox{$\bra{0}E_{123}\ket{0}\propto\bra{\txy}\sxy\rangle=0$} in the
fully symmetric phase and a similar computation in the dimer phase
yields  \mbox{$\bra{0}E_{123}\ket{0}\simeq\bra{\sy}E_{123}\ket{\sy}=0$}
as well. Thus there is no instability towards chiral symmetry breaking
in our approach either, and we need to address the dynamics of
$E_{123}$. Namely we compute the following dynamical correlation
\begin{equation}
\langle E_{123}(\omega)E_{123}\rangle
=\sum_q|\bra{q}E_{123}\ket{0}|^2\delta(\omega-\epsilon_q+\epsilon_0)
\label{eqn:dyn}
\end{equation}
where the sum runs over all eigenpairs $\{\epsilon_q,\ket{q}\}_q$ of the
impurity model~(\ref{eqn:HAIM}).  The above discussion shows that in the
fully symmetric phase, the lowest pole in Eqn.~\ref{eqn:dyn} is located
at the singlet gap energy $\omega_1=\epsilon_1-\epsilon_0$ with weight
\mbox{$Z_1=|\bra{1}E_{123}\ket{0}|^2\simeq(\sqrt{3}/4)^2|{}_\text{B}\bra{1}0\rangle_\text{B}|^2$}.

\begin{figure}[h]
\begin{center}
\resizebox{.5\textwidth}{!}{\includegraphics[viewport=35 80 780
600,clip]{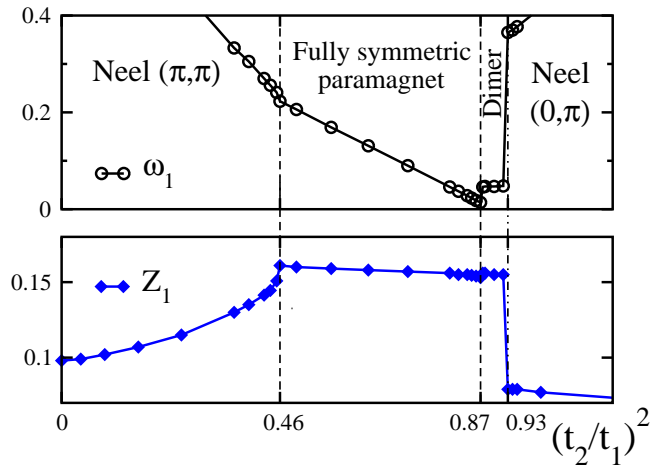}}
\end{center}
\caption{(color online) Evolution of the frequency $\omega_1$ and
integrated weight $Z_1$ of the lowest pole of the chiral correlation
\mbox{$\langle E_{123}(\omega)E_{123}\rangle$} with increasing
frustration.
%A local estimate of the spin gap is also shown in the paramagnetic
%phase for comparison (see text).
}
\label{fig:chi}
\end{figure}

In Figure~\ref{fig:chi} we show the evolution of $\omega_1$ and $Z_1$
with increasing \frust.
As expected, increasing the frustration $t_2/t_1$ from $t_2=0$ reduces
$\omega_1$ and increases $Z_1$, evidencing the enhancement of coherence
of the three spins on a plaquette. In both paramagnetic phases the
weight $Z_1$ decreases very slowly, in the $0.16-0.155$ range, showing
that \mbox{$|{}_\text{B}\bra{1}0\rangle_\text{B}|\simeq0.9$}. Hence, not
only are the bath and the impurity weakly entangled, but also the bath
states that accompany the lowest lying impurity states are quite
similar. In the following, this will allow us to ignore the bath
components of $\ket{0}$ and $\ket{1}$ temporarily. Although there is no
significant ``decoherence'' induced by the coupling of the impurity to
the bath, the bath does play a crucial role in determining the phase
diagram: without it we would not be describing a system in the
thermodynamic limit, able to sustain broken symmetry phases and
transitions between them.

The upper panel of Figure~\ref{fig:chi} shows that for
\mbox{$0.46\lsim(t_2/t_1)^2\lsim0.87$}, the singlet gap $\omega_1$
between $\ket{\txy}$ and $\ket{\sxy}$ decrease linearly towards zero as
\mbox{$(t_2/t_1)^2\to0.87$}. For comparison, the lowest pole of the
cluster dynamical spin-spin correlations \mbox{$\langle
m_{\bm{k}_\text{I}}(\omega)\cdot m_{-\bm{k}_\text{I}}\rangle$} is
obtained for \mbox{$\bm{k}_\text{I}=(\pi,\pi)$} in this region and
yields a local estimate of the spin gap of about $0.2\,t_1$.
%the order of .2 [WHICH UNITS],  estimated from estimated as the
%frequency of the lowest pole of the cluster dynamical correlation
%\mbox{$\langle m_{(\pi,\pi)}(\omega)\cdot m_{(\pi,\pi)}\rangle$}.

In the fully symmetric phase, any combination
\mbox{$\ket{\Psi}=a\ket{\txy}+b\ket{\sxy}$} is not an eigenstate and
instead has a dynamics with characteristic frequency $\omega_1$. As
\mbox{$(t_2/t_1)^2\to0.87$}, \mbox{$\omega_1\to0$} and all such states
become eigenstates with zero-energy. This obviously includes chiral
states, defined by \mbox{$\bra{\Psi}E_{123}\ket{\Psi}\neq0$}, and such
that \mbox{$ba^*-ab^*\neq0$}, but also the above-mentioned dimer states
$\ket{\sx}$ and $\ket{\sy}$, one of which will be eventually selected
upon increasing \frust above 0.87. 

%For the thermodynamic system this suggests that near the transition
%point, the abundance of low-lying singlets, appearing as low-energy
%poles in various correlation functions, results in many competing
%paramagnetic ground states.  
%
We now ask whether these chiral and dimer states exhaust the
possibilities of phases that can be described by plaquette DMFT and
which compete near \mbox{$(t_2/t_1)^2\simeq0.87$}. One route consists in
extending the above discussion of the chiral correlations. Namely we are
looking for any additional hermitian operator $A$ such that the
correlation $\langle A(\omega)A\rangle$ has its lowest pole at the
energy of the singlet gap $\omega_1$ in the fully symmetric phase.
According to Eqn.~\ref{eqn:dyn} this translates as
\mbox{$\bra{0}A\ket{0}\simeq\bra{\txy}A\ket{\txy}=0$} and
\mbox{$\bra{1}A\ket{0}\simeq\bra{\sxy}A\ket{\txy}\neq0$}. In such a
reduced subspace the possibilities are very limited: the set of
$2\times2$, hermitian, off-diagonal matrices is exactly spanned by the
two Pauli matrices $\sigma_x,\sigma_y$. In particular, the above
computation of $E_{123}$ can be recast in operator form as
\mbox{$E_{123}=\sqrt{3}/4\ \sigma_y$}. A similar computation for the
dimerization operator $D$ leads to \mbox{$D=\sqrt{3}/2\ \sigma_x$}. In
this synthetic form it is clear that we have exhausted the list of
linearly independant operators whose fluctuations become infinitely soft
at the transition point.

Interestingly, the local mean-field picture developed above allows us to
further elaborate on the nature of the chiral phase that competes with
the dimerized phase near \mbox{$(t_2/t_1)^2\simeq0.87$}. Indeed,
repeating the above computation for all triplets of impurity sites leads
to \mbox{$E_{123}=\sqrt{3}/4\ \sigma_y=-E_{234}=E_{341}=-E_{412}$}. 
%Upon taking the expectation value of these operator relations one
%obtains symmetry relations for the chiralities that are valid for {\it
%any} state $\ket{\Psi}$ in the zero-energy subspace, although they are
%non-trivial for chiral combinations only.  , so that
%$E_{123}E_{123}=(\sqrt{3}/4)^2=-E_{123}E_{234}=E_{123}E_{341}=-E_{123}E_{412}$,
%while the remaining correlations can be deduced by symmetry.
It is interesting to note the chiral spin liquid proposed by Wen in its
\mbox{large-$N$} treatment of the \jj model obeys the same spatial
symmetries.~\cite{wwz89} Further, Wen derived both i) a chiral liquid
and ii) a dimer solution of the mean-field equations.  Both solutions
were found to compete at large frustration $J_2\simeq J_1$, although
dimers are always energetically favored, in qualitative agreement with
the present study.

{\it Conclusion}
%
%We performed a plaquette DMFT study of the \mbox{$t_1-t_2-U$} model in
%its strong coupling regime at half-filling, paying special attention to
%possible symmetry-breaking instabilities.  We recovered the sequence of
%phases of the \jj model, with semi-classical \neel phases at small and
%large frustration, respectively, separated by a paramagetic region.
%Increasing \frust in this phase essentially {\it Strengths and
%weaknesses of the approach}
%
It is remarkable that a mean-field theory based on a plaquette is able
to describe the phases of the \jj model, as obtained within more exact
approaches.~\cite{ml05,nss08} The present study provides simple
caricatures of the local aspects of these phases.  Mean-field theory,
however, does not contain the effects of long-wavelength fluctuations
and is not expected to describe the location of the phase boundaries
accurately.  CDMFT also ignores more complicated forms of long-range
order which cannot be tiled with impurities. Overall we expect that,
upon increasing the size of the impurity, the phase boundaries in
Figure~\ref{fig:OP} will shift, moving the paramagnetic region
\mbox{$0.46\lsim(t_2/t_1)^2\lsim0.93$} closer to the values of the
nearby \jj model, namely \mbox{$0.4\lsim J_2/J_1\lsim0.6$}.~\cite{ml05}
\cite{nss08}.   Nevertheless, we expect the local physics described in
this paper to be robust and play an important role both at finite
temperature and finite doping, which deserves further study within
plaquette DMFT. In particular, doping this system in the region with
low-lying chiral fluctuations may generate orbital currents
or a $d$-density wave. Although no instability towards either of those
orders was detected in the $t_2=0$ case at finite doping,~\cite{mjm04}
the present study shows that this issue should be reconsidered for
$t_2\neq0$.

On the experimental side, the search for physical realizations of a \mbox{$t_1-t_2$} one-band
Hubbard model resulted in materials with either small $t_2/t_1$, such as
cuprates, which have \pp \neel$\!\!$-order at low-temperature, or very
large $t_2/t_1$, such as Li$_2$VOSiO$_4$, lying well in the \zp \neel
phase.~\cite{mcl00,mbp03} However, the recently synthetized material
PbVO$_3$~\cite{sct04}
% has $t_2/t_1$ in
%the $0.5-0.6$ range~\cite{tbs08,h08} and is therefore ideal to
has intermediate frustration~\cite{tbs08} and offers a possible
realization  of the  non-magnetic phase  described in this paper.  In
particular, Raman scattering under pressure  in various geometries
should be able to probe for the low-lying chiral
fluctuations~\cite{ss90,sfl91} and their evolution with frustration.

ACKNOWLEDGEMENT: This work was supported by the NSF under grant DMR No.
DMR 0528969.  Benoit Dou\c{c}ot, Olivier Parcollet, Marcello Civelli,
Claire Lhuillier, Gr\'egoire Misguich and Anne Passy are gratefuly
acknowledged for enlightening discussions.
%
%\bibliography{jc}
%\end{document}
%

%
\end{document}